\documentclass[11pt,a4paper]{article}
\usepackage{jheppub}
\usepackage{amsmath}
\usepackage{amssymb}
\usepackage{graphicx}
\usepackage[T1]{fontenc} 
\usepackage{booktabs}
\usepackage{slashed}
\usepackage[utf8]{inputenc}
\usepackage{xspace}
\usepackage{color}
\usepackage[capitalise, english]{cleveref}
\usepackage{siunitx}
\usepackage{soul}

\newcommand{\irrepbase}[2][0]{\ensuremath{\text{\boldirrep{#2}}}}

\newcommand{\boldirrep}{\textbf}
\newlength{\irrepwidth}
\newlength{\irrepbarthickness}
\newlength{\irrepbarheight}
\newcommand{\irrepbarbase}[1]{%
    \settoheight{\irrepbarheight}{\ensuremath{\text{\boldirrep{#1}}}}%
    \settowidth{\irrepwidth}{\ensuremath{\text{\boldirrep{#1}}}}%
    \makebox[0pt][l]{\ensuremath{\text{\boldirrep{#1}}}}%
    \rule[1.2\irrepbarheight]{\irrepwidth}{\irrepbarthickness}%
}

\def\primes#1#2{\count0=#1 \loop \ifnum\count0>0 \advance\count0 by -1 #2\repeat}
\newcommand{\irrep}[2][0]{\ensuremath{\irrepbase{#2}^{\primes{#1}{\prime}}}}
\newcommand{\irrepbar}[2][0]{\ensuremath{\irrepbarbase{#2}^{\primes{#1}{\prime}}}}

\newcommand{\GeV}{{\ensuremath{{\mathrm{GeV}}}}}

\definecolor{lightblue}{rgb}{1,.4,.5}

\definecolor{mkgreen}{rgb}{0.2,.70,.3}

\newcommand\vphiu[1]{v_{\Phi}^{u_{#1}}}
\newcommand\vphid[1]{v_{\Phi}^{d_{#1}}}
\newcommand\vphiusq[1]{(v_{\Phi}^{u_{#1}})^2}
\newcommand\vphidsq[1]{(v_{\Phi}^{d_{#1}})^2}

\newcommand\tanbeta{t_\beta}
\newcommand\tanbetau{t_{\beta_u}}
\newcommand\tanbetad{t_{\beta_d}}
\newcommand\tanbetaR{t_{\beta_R}}

\DeclareMathOperator{\tr}{Tr}

\def\lsim{\raise0.3ex\hbox{$\;<$\kern-0.75em\raise-1.1ex\hbox{$\sim\;$}}}
\def\gsim{\raise0.3ex\hbox{$\;>$\kern-0.75em\raise-1.1ex\hbox{$\sim\;$}}}

\newcommand{\AddrCERN}{%
 Theory Division, CERN, 1211 Geneva 23, Switzerland}

\newcommand{\AddrAHEP}{
  {\it AHEP Group, Instituto de F\'{\i}sica Corpuscular --
    C.S.I.C./Universitat de Val{\`e}ncia \\
    Edificio de Institutos de Paterna, Apartado 22085,
  E--46071 Val{\`e}ncia, Spain}}

\newcommand{\AddrWur}{%
Institut f\"ur Theoretische Physik und Astronomie, 
Universit\"at W\"urzburg\\
Emil-Hilb-Weg 22,
97074 Wuerzburg}

\newcommand{\AddrBonn}{%
Bethe Center for Theoretical Physics \& Physikalisches Institut der 
Universit\"at Bonn, \\
Nussallee 12, 53115 Bonn, Germany
}

\preprint{BONN-TH-2015-13\\ \hspace*{\fill} CERN-PH-TH-2015-279\\ \hspace*{\fill} IFIC/15-91}

\title{A constrained supersymmetric left-right model}

\author[a]{Martin Hirsch,} \emailAdd{mahirsch@ific.uv.es}
\affiliation[a]{\AddrAHEP}

\author[b,c]{Manuel E. Krauss,} \emailAdd{mkrauss@th.physik.uni-bonn.de}
\affiliation[b]{\AddrBonn}
\affiliation[c]{\AddrWur}

\author[b]{Toby Opferkuch,} \emailAdd{toby@th.physik.uni-bonn.de}

\author[c]{Werner Porod,} \emailAdd{porod@physik.uni-wuerzburg.de}

\author[d]{Florian Staub}\emailAdd{florian.staub@cern.ch}
\affiliation[d]{\AddrCERN}

\abstract{We present a supersymmetric left-right model which predicts gauge coupling unification close to the string scale and extra vector bosons at the TeV scale. The subtleties in constructing a model which is in agreement with the measured quark masses and mixing for such a low left-right breaking scale are discussed. It is shown that in the constrained version of this model radiative breaking of the gauge symmetries is possible and a SM-like Higgs is obtained. Additional CP-even scalars of a similar mass or even much lighter  are possible. The expected mass hierarchies for the supersymmetric states differ clearly from those of the constrained MSSM. In particular, the lightest down-type squark, which is a mixture of the sbottom and extra vector-like states, is always lighter than the stop. We also comment on the model's capability to explain current anomalies observed at the LHC.}
\begin{document}
\maketitle

\section{Introduction}
\label{sec:intro}

Supersymmetry (SUSY) has been the leading candidate for physics beyond the Standard Model (SM) for many years. Amongst many other appealing features, the minimal supersymmetric Standard Model (MSSM) predicts the existence of gauge coupling unification in the vicinity of the Planck scale \cite{Dimopoulos:1981yj,Ibanez:1981yh,Marciano:1981un,Einhorn:1981sx,Amaldi:1991cn,Langacker:1991an,Ellis:1990wk}. This has been considered as a clear hint
that SUSY is the next step towards a grand unified theory (GUT), see for instance Ref.~\cite{Mohapatra:1999vv} and references therein. 

In the past the Ansatz has often been made that all additional states beyond the MSSM, which are present in the GUT model,
are superheavy and subsequently do not play any role below the GUT scale. However, relaxing this condition and including either threshold effects of these states \cite{Kolda:1995iw,Baer:2000gf,Deppisch:2014aga},
or allowing some states to be ligher by orders of magnitude \cite{Drees:2008tc,Borzumati:2009hu,Esteves:2010si,Esteves:2011gk} could conceivably yield some insights into the GUT model even if TeV scale physics is still described by the MSSM, potentially extended by the Weinberg operator. However, the MSSM is facing increasing pressure over the last years: Minimal SUSY has problems explaining the size of the measured Higgs mass without losing attractiveness in terms of naturalness, see e.g. Ref.~\cite{Papucci:2011wy} and references therein, or being forced into regions with possibly dangerous charge- and colour-breaking minima of the scalar potential \cite{Camargo-Molina:2013sta,Camargo-Molina:2014pwa}.  Recently, the most minimal model, the constrained minimal supersymmetric standard model (CMSSM), has been ruled out at the \SI{95}{\%} confidence level \cite{Bechtle:2015nua}. Furthermore, minimal SUSY models require additional mechanisms in order to explain the measured neutrino mixings.

In the presence of additional gauge symmetries the Higgs mass can be increased at tree-level leading to
enhanced naturalness \cite{Batra:2003nj,Maloney:2004rc}. Additionally the presence of right-handed neutrinos, as predicted by $SO(10)$ GUT theories, allows for natural seesaw-like mechanisms \cite{Malinsky:2005bi}. A well motivated scenario which has not yet been studied is an $SO(10)$ GUT model which predicts left-right symmetry close to the TeV scale. 

Lately, left-right-symmetric theories have received increased interest. This is
due to the observation of anomalies in the \SI{8}{\TeV} LHC data occurring around \SI{2}{\TeV} \cite{Aad:2015owa,Khachatryan:2014dka,CMS:2015gla} as 
they can be interpreted as decays of a heavy $W'$ boson, see Refs. \cite{Dev:2015pga,Brehmer:2015cia,Deppisch:2015cua} and references therein. However, there has not yet been any attempt to embed these ideas in a top-down approach as it would be a natural candidate to originate from a GUT model.

There are many different realisations of left-right models proposed in
the literature. The most striking difference among different
left-right theories can be found in the sector that eventually breaks
the larger gauge group down to the SM gauge group.  The most appealing
choice would be the introduction of $SU(2)_R$ triplets which allow for
an automatic seesaw-mechanism of type I after left-right symmetry
breaking, see, e.g.,
Refs.~\cite{Cvetic:1983su,Kuchimanchi:1993jg,Babu:2008ep}. However,
besides being heavily constrained from vacuum stability arguments
\cite{Basso:2015pka}, the requirement of gauge coupling unification
usually requires the addition of extra intermediate supermultiplets
\cite{Majee:2007uv}.  In the presence of doublets, instead of
triplets, supersymmetric models consistent with gauge coupling
unification and a TeV-scale spectrum can be easily found
\cite{DeRomeri:2011ie,Arbelaez:2013hr}, while in the
non-supersymmetric case models with triplets are also possible
\cite{Arbelaez:2013nga}.  In the supersymmetric variants special care
has to be taken not to destroy the gauge coupling unification which already
works well in the MSSM.  The resulting conditions on the
particle content of the models, called the ``sliding scale'' mechanism in
these papers, have been discussed in Refs.~\cite{DeRomeri:2011ie,Arbelaez:2013hr}.

In this work, we present a left-right supersymmetric model consistent with gauge coupling unification and a minimal set of boundary conditions at the unification scale 
which maintains left-right symmetry down to energies accessible by the LHC without the
need of an intermediate scale. This paper is organised as follows: First, we discuss the basic model features and the
necessary conditions for successful gauge coupling unification as well as radiative symmetry breaking. We then present the
quark and lepton sectors in some detail. In section \ref{sec:pheno}, we address the Higgs mass and mixing as well as the expected squark hierarchies which differ from the CMSSM expectations. We close by commenting on two of the current LHC excesses.

\section{The Model}
\label{sec:model}

\subsection{Particle content, superpotential and gauge symmetry breaking}
We assume that $SO(10)$ is broken at the GUT scale and below this scale the remaining gauge group is left-right symmetric down to the SUSY scale, i.e. $\mathcal{G}=SU(3)_C\times SU(2)_L\times SU(2)_R\times U(1)_{B-L}$. The particle content of the model under consideration is given in \cref{tab:fieldcontent}.
\begin{table}
 \renewcommand\arraystretch{1.10}
\begin{center}
\begin{tabular}{c c c c c }
\toprule
Field & Multiplicity & $\mathcal{G}$ & \parbox{2cm}{$SO(10)$ Origin} & $\mathbb{Z}_2$ \\ 
\midrule
$Q$ & 3 & $(\irrep{3},\irrep{2},\irrep{1},+\frac{1}{3})$ & $\irrep{16}$ & -1 \\
$Q_c$ & 3 & $(\irrepbar{3},\irrep{1},\irrep{2},-\frac{1}{3})$ & $\irrep{16}$ & -1 \\
$L$ & 3 & $(\irrep{1},\irrep{2},\irrep{1},-1)$ & $\irrep{16}$ & -1 \\
$L_c$ & 3 & $(\irrep{1},\irrep{1},\irrep{2},+1)$ & $\irrep{16}$ & -1 \\
$S$ & 3 &  $(\irrep{1},\irrep{1},\irrep{1},0)$ & $\irrep{1}$ & -1\\
$\delta_d$  & 1 & $(\irrep{3},\irrep{1},\irrep{1},-\frac{2}{3})$ & $\irrep{10}$ & -1 \\
$\bar{\delta}_d$ & 1 & $(\irrepbar{3},\irrep{1},\irrep{1},+\frac{2}{3})$ & $\irrep{10}$ & -1 \\
$\Psi$, $\Psi_c$ & 2 & $(\irrep{1},\irrep{1},\irrep{1}, \pm2)$ & $\irrep{120}$& -1 \\
\hline
$\Phi$ & 2 & $(\irrep{1},\irrep{2},\irrep{2},0)$ & $\irrep{10}$, $\irrep{120}$ & 1 \\
$\chi_c$, $\bar{\chi}_c$ & 1 & $(\irrep{1},\irrep{1},\irrep{2}, \mp1)$ & $\irrepbar{16}$, $\irrep{16}$ & 1\\
\bottomrule
\end{tabular}
\end{center}
\caption{The matter sector and Higgs sector field content of the supersymetric left-right model. Generation indices have been suppressed and the index $c$ refers to the equivalent SM field which transforms under $SU(2)_R$. The gauge group is such that $\mathcal{G}=SU(3)_C\times SU(2)_L\times SU(2)_R\times U(1)_{B-L}$. Note that We also assume the usual matter parity.}
\label{tab:fieldcontent}
\end{table} 
Here, $\Phi$ is a bi-doublet which comes in two generations
\begin{align}
\Phi^a &= \begin{pmatrix} H_d^{a0} & H_u^{a+} \\ H_d^{a-} & H_u^{a0} \end{pmatrix}.
\end{align}
The SM-like Higgs will be in general a superposition of the four neutral components of these bi-doublets. The conventions for 
the fields which will be responsible for the breaking of $SU(2)_R \times U(1)_{B-L}$ are:
\begin{align}
\chi_c&= \begin{pmatrix} \chi_c^0 \\ -\chi_c^- \end{pmatrix}, \quad \quad  \bar{\chi}_c = \begin{pmatrix} \bar{\chi}^+_c \\ -\bar{\chi}_c^0 \end{pmatrix}
\end{align}
Using this field content the renormalizable superpotential allowed under both the gauge symmetries $\mathcal{G}$ and matter parity \cite{Dev:2009aw,BhupalDev:2010he} is
\begin{align}\label{eq:superpotential}
W &= Y_{Q_a} Q \,\Phi^a Q_c + Y_{L_a} L\, \Phi^a L_c + Y_{\delta_d} Q_c \bar{\chi}_c \delta_d +  Y_S L_c \chi_c S + Y_\Psi L_c \bar{\chi_c} \Psi_c \\
&+ \frac{\mu_S}{2} S^2 + \mu_\Phi^{ab} \Phi_a \Phi_b + \mu_{\chi_c} \bar{\chi}_c \chi_c + M_\delta \delta_d \bar{\delta}_d  + M_\Psi \Psi \Psi_c \,. \notag
\end{align}
Here, all generation, $SU(3)$ and $SU(2)$ indices are suppressed. 

Spontaneous symmetry breaking occurs when the neutral components of $\Phi$ and $\chi$-fields receive vacuum expectation values (VEVs)
\begin{subequations}
\begin{align}
H_d^{a0} &= \frac{1}{\sqrt{2}}\left( \sigma_d^a +i\varphi_d^a +\vphid{a}\right)\, , \label{eq:decomp_Hd}\\
H_u^{a0} &= \frac{1}{\sqrt{2}}\left(\sigma^a_u+i\varphi^a_u+\vphiu{a}\right)\,,\label{eq:decomp_Hu} \\
\chi_c^0 &= \frac{1}{\sqrt{2}} \left(\sigma_{\chi_c}+i \varphi_{\chi_c}+v_{\chi_c}\right)\,,\label{eq:decomp_Chic}\\
\bar{\chi}^0_c&=\frac{1}{\sqrt{2}}\left(\bar{\sigma}_{\bar{\chi}_c}+i\bar{\varphi}_{\bar{\chi}_c}+v_{\bar{\chi}_c}\right) \,. \label{eq:decomp_Chibarc}
\end{align}
\end{subequations}
We make use of the following definitions of the VEVs
\begin{subequations}
\begin{align}
v_R^2 &= v_{\chi_c}^2+v_{\bar{\chi}_c}^2\,,\\
v_L^2&= \vphidsq{1}+\vphidsq{2}+\vphiusq{1}+\vphiusq{2}\,,
\end{align}
\end{subequations}
where we use three angles to parametrise the VEVs
\begin{subequations}
\begin{align}
\vphiu{1} &= v_L \sin\beta \sin\beta_u, \quad\quad \vphid{1} = v_L \cos\beta \sin\beta_d,\\
\vphiu{2} &= v_L \sin\beta \cos\beta_u, \quad\quad \vphid{2} = v_L \cos\beta \cos\beta_d.
\end{align}
\end{subequations}
In this parametrisation $v_L$ is the electroweak VEV and $\beta$ is the usual mixing angle projecting out the $SU(2)_L$ would-be-Goldstone bosons as in the MSSM. In general, the gauge symmetries are broken in two steps 
\begin{align}\label{eq:breakingchain}
\begin{split}
\mathcal{G}&=SU(3)_C \times SU(2)_L \times SU(2)_R \times U(1)_{B-L},\\
&\xrightarrow{v_R} SU(3)_C \times SU(2)_L \times U(1)_Y, \\
&\xrightarrow{v_L} SU(3)_C \times U(1)_{\text{EM}}=\mathcal{G}_{SM}.
\end{split}
\end{align}
However, if $v_R$ does not exceed the $\SI{}{\TeV}$ range, one can assume to a good approximation a one-step breaking $\mathcal{G} \to \mathcal{G}_{SM}$
which also occurs close to the SUSY breaking scale $M_{\rm SUSY}=\sqrt{m_{\tilde t_1} m_{\tilde t_2}}$, where $\tilde t_1$ and $\tilde t_2$ are the two mostly stop-like up-type squarks.

\begin{figure}
\centering
  \includegraphics{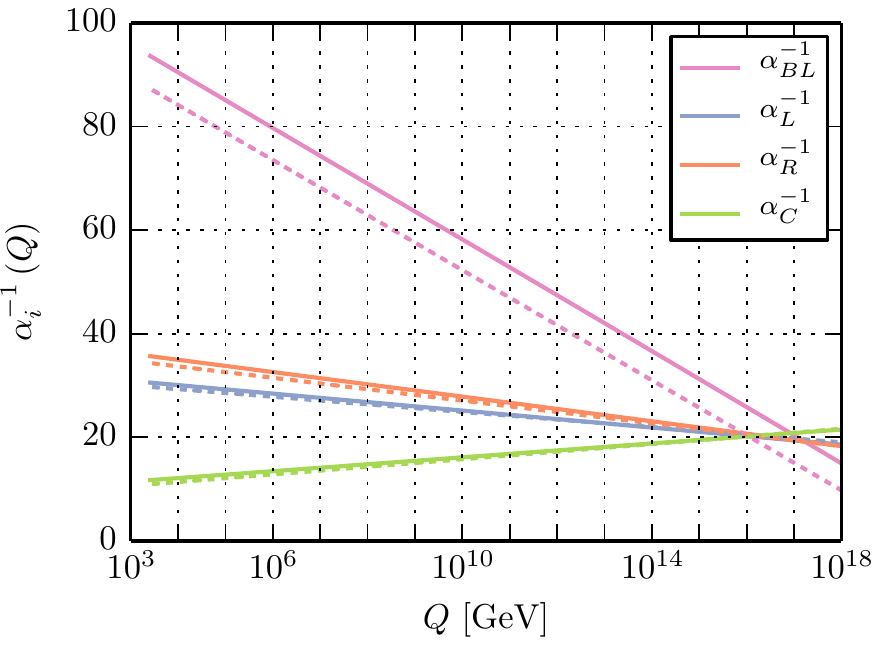}
\caption{Running of the gauge couplings at one-loop (dashed lines) and two-loop (solid lines) in the left-right phase of the model. The two-loop results includes the one-loop threshold corrections arising at both the electroweak and SUSY scales. Additionally the running of the couplings is shown from the SUSY scale rather than the $SU(2)_R$ breaking scale $v_R$. In this figure the GUT normalised $g_{BL}$ is plotted. The normalisation is given by $g_{BL}=\sqrt{\frac{3}{2}} g_{BL}^{\text{GUT}}$. The parameters for the two-loop running are $\tanbeta=5$, $\tanbetau=6.5$, $\tanbetaR=0.85$, $m_0=\SI{1.5}{\TeV}$, $M_{1/2}=\SI{750}{\GeV}$, $A_0=\SI{1}{\TeV}$, $Y_{\delta_d}^i=0.15$ and $v_R=6~$TeV.
}
\label{fig:GCU}
\end{figure}

We show in \cref{fig:GCU} the running of the gauge couplings in this model. This shows, the assumption of a left-right breaking close to the $\SI{}{\TeV}$
scale is consistent with gauge coupling unification. We find that the unification scale is significantly larger than in the MSSM, lying in the range (1--$4)\times\SI{E17}{\GeV}$. The increased scale of unification arises for two reasons. Firstly, the one-loop threshold corrections are large. 
This is due to the mass spectrum being spread over several \SI{}{\TeV} leading to large logarithms in the threshold corrections. Secondly, the beta coefficient of the $U(1)_{B-L}$ gauge coupling is large, taking the value $29/2$. Consequently the unification scale becomes extremely sensitive on the initial value of $g_{BL}$, which also generically receives large corrections due to the thresholds.  
Subsequently, maintaining gauge coupling unification requires that the mass spectrum of the theory remain as light as possible, leading to the prediction of a small $SU(2)_R$ breaking scale. Finally, the running values of the new couplings at $M_{\rm SUSY}$ are $g_{BL} \simeq 0.44$ and $g_R \simeq 0.59$. 

As a consequence of the symmetry breaking, there are two additional massive gauge boson. Their masses can be approximated by
\begin{subequations}
\begin{align}
M_{Z^\prime}^2&\simeq \frac{1}{4} \left(\left(g_{BL}^2+g_R^2\right) v_R^2 + \frac{g_R^4}{(g_{BL}^2+g_R^2)} v_L^2\right)\,, \\
M_{W^{'\pm}}^2 &\simeq \frac{1}{4} g_R^2 \left(v_L^2+v_R^2\right)\,.
\end{align}
\end{subequations}
Beside the Weinberg angle, two additional rotation angles for the neutral gauge bosons are required, 
while only one extra angle is required for the charged gauge bosons. 
The mixing angles between the mass eigenstates of the new gauge bosons are given by
\begin{align} \label{eq:zzp_mixing}
 \sin2\Theta_{ZZ'} &\simeq \frac{2 g_R^2 v_L^2 \sqrt{g_{BL}^2 g_R^2 +g_L^2 (g_{BL}^2+g_R^2)}}{(g_{BL}^2+g_R^2)^2 v_R^2}\,, \\
 \tan 2\Theta_{WW'} &=\frac{4 g_L g_R v_L^2}{\left(g_R^2(v_L^2+v_R^2)-g_L^2 v_L^2\right)} \frac{\tanbeta (1+\tanbetad\tanbetau)}{(1+\tanbeta^2)\sqrt{(1+\tanbetad^2)(1+\tanbetau^2)}}\,,
 \label{eq:wwp_mixing}
 \end{align}
where we have used the abbreviations $\tanbeta=\tan\beta$, $\tanbetau=\tan\beta_u$ and $\tanbetad=\tan\beta_d$. 

There are a number of choices for which parameters to solve the six minimisation conditions for the vacuum. Here we solve for the following parameters 
\begin{align}
&\{(\mu_\Phi^{(1,1)})^2,B_{\mu_\Phi}^{(1,1)},B_{\mu_\Phi}^{(1,2)},B_{\mu_\Phi}^{(2,2)},\mu_{\chi_c}, B_{\mu_{\chi_c}}\}. 
\end{align}
This set of parameters allows for unified soft-masses at the GUT scale while also allowing the minimisation conditions to be solved analytically. 
The second advantage of this set of parameters is that both $\mu_{\chi_c}$ and $B_{\mu_{\chi_c}}$ appear in only two of the six tadpole equations and can therefore be solved independently of the remaining parameters. We obtain
\begin{subequations}\label{eq:SU2Rtadpoleeqns}
\begin{align}\label{eq:SU2Rtadpoleeqn1}
|\mu_{\chi_c}|^2&=\frac{1}{2} \bigg( \left(\Delta m_{\chi_c}^2-\frac{1}{4} g_R^2 v_L^2 \cos2\beta\right) \frac{1+\tanbetaR^2}{1-\tanbetaR^2} +\sum m_{\chi_c}^2 -\frac{1}{4}\left(g_{BL}^2+g_R^2\right)v_R^2 \bigg)\,,  \\
 &\simeq  \frac{1}{1-\tanbetaR^2} \bigg(m^2_{\chi_c} \tanbetaR^2 - m^2_{\bar \chi_c}  \bigg) - \frac{1}{2} M^2_{Z'}\,,\label{eq:SU2Rtadpoleeqn1b}\\
B_{\mu_{\chi_c}}&=\frac{1}{2} \bigg( \left(-\Delta m_{\chi_c}^2 +\frac{1}{4}g_R^2 v_L^2 \cos2\beta \right) \frac{2 \tanbetaR}{1-\tanbetaR^2}+\frac{1}{4}\left(g_{BL}^2+g_R^2\right)v_R^2 \frac{2 \tanbetaR}{1+\tanbetaR^2} \bigg)\,, 
\end{align}
\end{subequations}
where $\tanbetaR = \tan\beta_R=v_{\chi_c}/v_{\bar \chi_c}$, $\Delta m_{\chi_c}^2=m_{\chi_c}^2-m_{\bar{\chi}_c}^2$ and $\sum m_{\chi_c}^2=-\left(m_{\chi_c}^2+m_{\bar{\chi}_c}^2\right)$. 
We assume that SUSY breaking in the visible sector is triggered by gravity and therefore make use of mSugra-like boundary conditions at the GUT scale, i.e.
subsequently we impose the unification of the following soft-parameters:
\begin{subequations}
\begin{align}
m_0^2 \delta_{ij} &=m_Q^2 \delta_{ij} = m_{Q_c}^2 \delta_{ij}= m_L^2  \delta_{ij} = m_{L_c}^2  \delta_{ij}= m_S^2\delta_{ij}  \notag \\ 
&=m_{\Psi}^2 \delta_{ij}=m_{\Psi_c}^2 \delta_{ij}= m_{\bar{\delta}_d}^2= m_{\delta_d}^2m_\Phi^2\delta_{ij} = m_{\chi}^2 = m_{\chi_c}^2 \,, \\
M_{1/2} &= M_{B-L} = M_R = M_L = M_3 \,.
\end{align}
\end{subequations}
The trilinear soft-breaking couplings are related to the superpotential couplings by an univeral parameter $A_0$
\begin{align}
T_i &= A_0 Y_i\,, \qquad i=Q,L,\delta_d,\Psi,S\,.
\end{align}
The resulting free parameters at the GUT scale that are of interest for phenomenological studies\footnote{We consider here the vector-like leptons $\Psi$,$\Psi_c$ and their scalar superpartners as spectator fields only necessary for gauge coupling unification. As such in all numerical studies we chose $M_\Psi = \SI{1}{\TeV}$ and set the corresponding $B_{\mu_\Psi}$ term to zero. Relaxing this assumption could have interesting consequences for collider phenomenology as well as flavour observables \cite{Falkowski:2013jya}.} are $m_0, M_{1/2}, A_0, \tanbeta, \tanbetau, \tanbetaR, (\mu_{\Phi}^{(2,2)})^2, Y_{\delta_d} Y_S$, $Y_\Psi$ and $M_\delta$.

\begin{figure}
\centering
  \includegraphics[width=\linewidth]{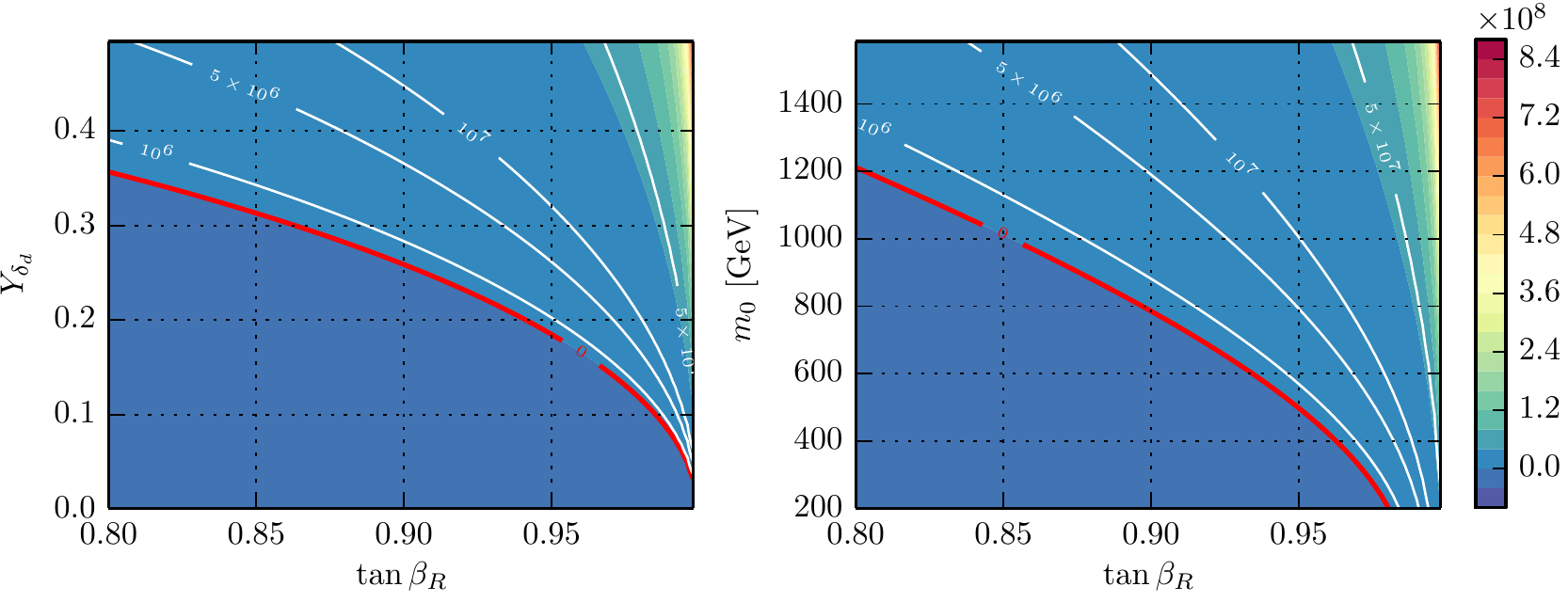}
\caption{Illustration of the constraints on the parameter space arising through the requirement of consistent solutions of the tadpole equations. The figures show contours of the $\mu_{\chi_c}^2$ values as function of either, $Y_{\delta_d}$ and $\tanbetaR$ (left) or $m_0$ and $\tanbetaR$ (right). Here, we have chosen the couplings $Y_S$ and $Y_\psi$ to be zero in order to reduce the dimensionality of the parameter space, for a detailed discussion of the effect of these parameters see the text. The red contour corresponds to where $\mu_{\chi_c}^2=0$, therefore the parameter space underneath this contour is excluded as $\mu_{\chi_c}^2<0$ in this region. The parameter values chosen correspond to left: $v_R= \SI{7}{\TeV}$, $m_0=\SI{750}{\GeV}$ and right: $v_R=\SI{7}{\TeV}$, $Y_{\delta_d}=0.25$. Other parameter values are, $\tanbeta =10$, $A_0=\SI{500}{\GeV}$ and $M_{1/2}=\SI{1}{\TeV}$.}
\label{fig:tadpole_plots}
\end{figure}

Using these boundary conditions, the running of the soft masses appearing in \cref{eq:SU2Rtadpoleeqns} can be approximated analytically at the one-loop level. This yields the results
\begin{subequations}\label{eq:masssplittingrunning2}
\begin{align}\label{eq:masssplittingrunning2a}
\Delta m_{\chi_c}^2 &\simeq \frac{1}{4\pi^2}\bigg((A_0^2 + 3 m_0^2)\left[3Y_{\delta_d}^\dag Y_{\delta_d} - \tr Y_S^\dag Y_S + \tr Y_\psi^\dag Y_\psi\right]\bigg) \ln \left(\frac{M_{\rm GUT}}{M_{\rm SUSY}}\right)\, , \\
\sum m_{\chi_c}^2&\simeq-2m_0^2+\frac{1}{4\pi^2} \bigg(\left(A_0^2 + 3 m_0^2\right)\left[3Y_{\delta_d}^\dag Y_{\delta_d} + \tr Y_S^\dag Y_S + \tr Y_\psi^\dag Y_\psi\right]\notag\\
&\qquad\qquad\qquad  -\left(3 g_{BL}^2+6g_R^2\right) M_{1/2}^2 \bigg) \ln \left(\frac{M_{\rm GUT}}{M_{\rm SUSY}}\right)\,.
\end{align}
\end{subequations}
In order to obtain spontaneous symmetry breaking one requires $\mu_{\chi_c}^2 \geq 0$, namely the RHS of \cref{eq:SU2Rtadpoleeqn1} must be greater than or equal to zero. This constraint excludes an area of the parameter space as a function of the couplings $Y_{\delta_d},$ $Y_\psi$, $Y_S$, the soft-breaking parameters $m_0$, $A_0$ and the $SU(2)_R$ VEV $v_R$. 
As the large $SU(2)_R$ $D$-terms in \cref{eq:SU2Rtadpoleeqn1} add negatively to $|\mu_{\chi_c}^2|$, the contribution from the soft masses has to account for the positivity requirement. From  \cref{eq:masssplittingrunning2a} one sees that $m^2_{\chi_c}>m^2_{\bar \chi_c}$ as long as $\Delta Y^2 \equiv 3 {\rm Tr}\, Y_{\delta_d}^\dagger Y_{\delta_d}+{\rm Tr}\, Y_{\psi}^\dagger Y_{\psi}^2 - {\rm Tr}\, Y_S^\dagger Y_S>0$, so that \cref{eq:SU2Rtadpoleeqn1b}  requires $\tanbetaR$ close to, but smaller than one. Values of $\tanbetaR$ significantly smaller than unity require a large splitting $\Delta m^2_{\chi_c} $, which can be achieved by increasing $m_0,~A_0$ or $
\Delta Y^2$. We exemplify this behaviour in \cref{fig:tadpole_plots} where we show the 
contours of different $\mu^2_{\chi_c}$ values as functions of $\tanbetaR$, $m_0$ and $Y_{\delta_d}$,\footnote{The approximations applied in \cref{fig:tadpole_plots} do not include the running of $Y_{\delta_d}$. Generically, this running increases the size of the couplings, but does not qualitatively modify the behaviour shown in the figure.} highlighting the $|\mu_{\chi_c}|^2=0$ contour in red. 
\subsection{Running of first and second generation Sfermions and gauginos}
In the CMSSM, which contains similar boundary conditions, one can obtain simple expressions at the one-loop level for the first and second generation sfermion soft-masses relating their size at $M_{\rm SUSY}$ to the high-scale parameters $m_0$ and $M_{1/2}$ \cite{Martin:1997ns}:
\begin{subequations}
\begin{align}
m_{q}^2(M_{\rm SUSY}) &\simeq  m_0^2 + 5.2 M_{1/2} \\
m_{d}^2(M_{\rm SUSY}) &\simeq  m_0^2 + 4.8 M_{1/2} \\
m_{u}^2(M_{\rm SUSY}) &\simeq  m_0^2 + 4.8 M_{1/2} \\
m_{l}^2(M_{\rm SUSY}) &\simeq  m_0^2 + 0.50 M_{1/2} \\
m_{e}^2(M_{\rm SUSY}) &\simeq  m_0^2 + 0.15 M_{1/2} 
\end{align}
\end{subequations}
Using the same Ansatz in our model, assuming $M_{\rm SUSY} \simeq v_R$, we obtain:
\begin{subequations}
\begin{align}
m_{Q}^2 (M_{\rm SUSY}) &\simeq   m_0^2 + 3.6 M_{1/2} \\
m_{Q^c}^2 (M_{\rm SUSY}) &\simeq   m_0^2 + 3.5 M_{1/2} \\
m_{L}^2 (M_{\rm SUSY}) &\simeq   m_0^2 + 0.44 M_{1/2} \\
m_{L^c}^2 (M_{\rm SUSY}) &\simeq   m_0^2 + 0.36 M_{1/2} 
\end{align}
\end{subequations}
Even if one must bear in mind that the correct coefficients get modified at the two-loop level, one can already see two main differences: (i)  the mass splitting between left- and right-sleptons  and squarks respectively is comparatively smaller than in the CMSSM, (ii) the squark masses don't grow so rapidly with increasing $M_{1/2}$ as they do in the CMSSM.

For the running gaugino masses one can obtain a rough estimate of the expectations of the CMSSM against this model at one-loop using the relation $M_a = g^2_a/g_{GUT}^2 M_{1/2}$. 
In the CMSSM, one obtains 
\begin{eqnarray}
M_1 \simeq 0.4 M_{1/2}\,,\, M_2 \simeq 0.75 M_{1/2}\,,\, M_3 \simeq 2.15 M_{1/2}\,,\quad
\end{eqnarray}
while this model predicts 
\begin{eqnarray}
&M_{B-L} \simeq 0.5 M_{1/2}\,,\,\,M_{L} \simeq 0.7 M_{1/2}\,,\,\,M_{R} \simeq 0.6 M_{1/2}\,,\,\, M_{3} \simeq 1.85 M_{1/2}\,.
\end{eqnarray}
Thus, the lightest gaugino is the one of the $U(1)_{B-L}$ gauge group. Moreover, the gluino is also lighter for the same value of $M_{1/2}$ as in the CMSSM despite the increased GUT scale.
\subsection{Quark masses and mixing}
In the simplest left-right model with only one generation of bi-doublets and no vector-like quarks, the quark mixing is trivial and the 
CKM matrix can't be generated. An Ansatz often used in literature to cure this problem is to add vector-like quarks which generate 
the CKM matrix via the mixing with the SM quarks. In the case of vector-like states  which mix with the d-quarks  and only one generation of bi-doublets, 
the two mass matrices read
\begin{equation}\label{eq:MuMd}
M_u = \frac{v_u}{\sqrt{2}} Y \,,\quad M_d=\begin{pmatrix}   
                                              \frac{v_d}{\sqrt{2}} Y & \frac{v_{\bar \chi_c}}{\sqrt{2}}  Y_{\delta_d} \\
                                              \tilde{m}  & M_{\delta_d} 
                                             \end{pmatrix}\,.
\end{equation}
To be very general, we kept a term $\tilde{m}$ which is actually absent in our model. 
$M_u$ is diagonalised by two $3 \times 3$ matrices $U^R_u$, $U^L_d$, and $M_d$ by $4 \times 4 $ matrices $U^R_d$, $U^L_d$. The measured CKM matrix  $V_{\rm CKM}$
must be reproduced by the $3 \times 3$ block related to the usual SM-quarks
of the matrix
\begin{equation}
V_{\rm CKM}^{4\times4} = \tilde U^u_{L} (U^d_{L})^\dagger \,,
\end{equation}
where $\tilde U^u$ is  $U^u$ enlarged artificially by a row and column of zeros 
apart from the (4,4) entry which is set to 1. One can always assume a basis, where $U^u$ is diagonal and the entire quark mixing is encoded in 
$U^d_L$. 
In this case and for $M_{\delta_d} \gg m_b$, one finds the seesaw condition 
\begin{equation}
\label{eq:seesaw}
M  =  \frac{v^2_{\bar \chi_c} Y_{\delta_d} Y_{\delta_d}^\dagger}{2} - \tilde{y} \tilde{y}^\dagger \,,
\end{equation}
with 
\begin{subequations}
\begin{align}
M&= V^*_{\rm CKM} \text{diag}(m^2_d, m_s^2, m_b^2)V^T_{\rm CKM}- \frac{v_d^2 Y_{Q} Y_Q^\dagger}{2}\,,\\
 \tilde{y} &= \frac{v_u Y_Q \tilde{m}^\dag + M_{\delta_d}^* v_{\bar\chi_c}Y_{\delta_d}}{\sqrt{2 (|\tilde{m}|^2 - M_{\delta_d}^2)}}\,.
\end{align}
\end{subequations}
Using  $\text{det}(A + u v^T) = (1+v^T A^{-1} u)\text{det}(A)$ for an invertible matrix $A$ and vectors $u$,$v$, one finds that the determinant of the RHS of \cref{eq:seesaw} always vanishes. This observation together with the 
extension of the same lemma where $u,v$ are $n \times m$ matrices yields the condition
\begin{equation}
1 -   \frac{v_d^2}{2} Y_Q \left(V^*_{\rm CKM} \text{diag}(m^2_d, m_s^2, m_b^2)V^T_{\rm CKM}\right)^{-1} Y_Q^\dagger = 0\,,
\end{equation}
for the LHS. Keeping in mind that $Y_Q$ is diagonal in the chosen basis, we finally find
\begin{equation}
  \frac{v_d^2}{2} Y_Q^2  = V^*_{\rm CKM} \text{diag}(m^{2}_d, m_s^{2}, m_b^{2})V^T_{\rm CKM}\,.
\end{equation}
Thus, there is only a solution to \cref{eq:seesaw} if quark mixing vanishes, otherwise the system is over constrained. We checked numerically that this conclusion holds also independently of the seesaw matrix and that the inclusion of radiative corrections does not alleviate this problem if one demands that all interactions are perturbative. Therefore, the best way to incorporate correct quark mixing in left-right models is to include a second generation of bi-doublets. However, the vector-like quarks in this model play still a crucial role because they are needed for radiative symmetry breaking as discussed below. \\
In the presence of two generations of bi-doublets, the Yukawa coupling in the left-right phase is related to the usual up- and down-type MSSM Yukawas $Y_u$, $Y_d$ via
\begin{subequations}
\begin{align}\label{eq:yukawadef1}
	Y_{Q_1}&=-\frac{Y_d\sqrt{1+\tanbetad^2}  -Y_u\sqrt{1+ \tanbetau^2} }{\tanbetad-\tanbetau}&&\xrightarrow{\tanbetad=0} &Y_{Q_1}&=\frac{Y_d-Y_u\sqrt{1+ \tanbetau^2}}{\tanbetau}\,,\\
	Y_{Q_2}&=\frac{\tanbetau Y_d\sqrt{1+\tanbetad^2} -Y_u\tanbetad\sqrt{1+ \tanbetau^2}}{\tanbetad-\tanbetau}&&\xrightarrow{\tanbetad=0} &Y_{Q_2}&=-Y_d\,.\label{eq:yukawadef2}
\end{align}
\end{subequations}
To keep $Y_{Q_2}^{(3,3)}$ perturbative up to the GUT scale, either $\tanbetau$ or $\tanbetad$ is restricted to very small values. Therefore we choose to always take $\tanbetad=0$. 

Our Ansatz to calculate $Y_{Q}$ numerically is as follows: we derive values for $Y_d$ and $Y_u$ to reproduce the known CKM matrix and quark masses. Here, two difficulties have to be taken into account: (i) the mixing with the vector-like quarks which is inevitable because we need a non-vanishing $Y_{\delta_d}$, and (ii) the full one-loop radiative corrections to all quarks. From the obtained values of $Y_d$ and $Y_u$, $Y_Q$ is calculated. Since $Y_Q$ affects the one-loop corrections to the quarks entering the calculation of $Y_d$ and $Y_u$, this procedure is iterated until a convergence has been reached.

We now briefly comment on the constraints arising from introducing vector-like quarks. 
Firstly, let us consider the constraints arising from quark flavour observables due to mixing between the vector-like and down-type quarks. 
The key point to note is that the introduced vector-like quarks only mix with the right-handed SM 
quarks due to the superpotential term $Y_{\delta_d} Q_c \bar{\chi}_c \delta_d$. The 
strongest bound stems from the kaon mixing where one also has to include the mixing of
heavy vector bosons which scale as $v_L^2/v_R^2$, see \cref{eq:zzp_mixing}
and (\ref{eq:wwp_mixing}). Recent collider data requires that the $W^\prime$ mass be at least approximately $\SI{2}{\TeV}$ \cite{ATLAS:2015nsi}. Apart from that, it has also been shown that kaon mixing constraints require the $W^\prime$ boson in left-right models to be at least approximately $\SI{3}{\TeV}$ in the non-supersymmetric case, \cite{Bertolini:2014sua} and at least $\SI{2}{\TeV}$ in supersymmetric models due to gluino contributions \cite{Zhang:2007qma}. Both of these bounds must be recast for the specific model in question; however, they do not change the conclusion that both the $W-W^\prime$ and $Z-Z^\prime$ mixing should be highly suppressed.
As discussed in \cref{sec:diboson}, the vector boson mixing is at
most $10^{-3}$ in our model. The mixing in the right-handed $d$-quark sector is at most 
$m_b/M_{\delta} \lsim 10^{-2}$. In the kaon mixing, both the squares of the quark and vector
boson mixing enters, implying that we can easily avoid this bound.


Lastly, one must consider the impact of the vector-like quarks on the electroweak 
precision observables. Due to the tree-level coupling of the vector-like quarks to $Z$-bosons, there will in general be a non-negligible contribution. The corresponding bounds have
been obtained in Ref.~\cite{Aguilar-Saavedra:2013qpa}: while the masses of the vector-like quarks should be
$\gsim \SI{600}{\GeV}$, the mixing with the SM quarks is constrained to $ | V_{\text{CKM},34}^{4\times 4}| \lsim 0.04$.\footnote{Interestingly, the bounds from the hadronic ratio $R_b$ are stronger than those arising from the oblique parameters for the considered case of down-type vector-like quarks.}

\subsection{Lepton sector}
\label{subsec:lepton}

In the lepton sector, we find equivalent relations between $Y_{L_1}$, $Y_{L_2}$ and both the lepton Yukawa coupling $Y_e$ as well as neutrino Yukawa coupling $Y_\nu$ as 
\cref{eq:yukawadef1,eq:yukawadef2} for the quark sector. Because of the additional gauge singlet $S$ as well as the two generations of extra vector-like leptons $\Psi, \Psi_c$, there are more free parameters in
the lepton sector as in the quark sector. Thus, the calculation of $Y_e$ and $Y_\nu \equiv \sqrt{2} m_D/v_u$ is in general more complicated.
In the limit $v_{\bar \chi_c} Y_\Psi  M_\Psi^{-1} \to 0$, the 
SM charged leptons decouple from the vector-like states and correspondingly, $v_d Y_e = -\sum_a v_\Phi^{d_a} Y_{L_a} $ can be diagonalized as usual, which fixes one linear combination of $Y_{L_1}$ and $Y_{L_2}$. The other necessary combination of $Y_{L_1}$ and $Y_{L_2}$ can be obtained from neutrino data. 

The neutrino masses can be calculated in the the seesaw approximation, 
which give the following expressions for light (heavy) neutrinos \cite{Forero:2011pc}:
\begin{align}
m_\nu^{\rm light} &\simeq \frac{2}{v_{\chi_c}^2} m_D\, (Y_S^T)^{-1} \mu_S Y_S^{-1} m_D^{T} \,, \nonumber \\ 
m_{\nu_h} &\simeq \frac{v_{\chi_c}}{\sqrt{2}} Y_S\,.
\end{align}
While the light neutrinos are Majorana states, the six heavy states form three quasi-Dirac pairs.

Since the right-handed neutrinos are part of the $L_c$ doublets, 
it is in general not possible to simultaneously diagonalize $Y_e$ and $Y_S$, as opposed to inverse seesaw models with the SM gauge group or with $U(1)_R\times U(1)_{B-L}$. 
However, one can always choose a basis with diagonal $Y_e$, $\mu_S$ and $M_\Psi$.  Therefore, the PMNS matrix can be fitted by the linear combination $
- \sum_a \frac{1}{\sqrt{2}} v_\Phi^{u_a} Y_{L_a} \equiv m_D$. Alternatively, one can work with diagonal $m_D$ and use
$Y_S$ to fit neutrino data, or allow off-diagonals in both terms.


\section{Phenomenology}
\label{sec:pheno}
In this section we discuss various phenomenological features of the model, focusing on aspects of the mass spectrum that differ compared to the MSSM, as well as on current excesses reported by the LHC experiments. A discussion of the rich flavour phenomenology of this model which provides several new sources for lepton and quark flavour violation, as well as of the dark matter scenarios is beyond the scope if this work and will be given elsewhere.

The numerical results of the model have been calculated using \texttt{SPheno} \cite{Porod:2003um,Porod:2011nf}, while the implementation of the model into \texttt{SPheno} was performed using the mathematica code \texttt{SARAH} \cite{Staub:2008uz,Staub:2009bi,Staub:2010jh,Staub:2012pb,Staub:2013tta,Staub:2015kfa}. This allows one to calculate the full one-loop spectrum as well as the dominant two-loop contributions to the CP-even Higgs masses \cite{Goodsell:2014bna,Goodsell:2015ira}. 
\subsection{Higgs sector}
\label{subsec:Higgs}
After the would-be Goldstone bosons are rotated out, the Higgs sector comprises six neutral $CP$-even states ($\sigma_i$, see Eqs.~(\ref{eq:decomp_Hd} -- \ref{eq:decomp_Chibarc})), four neutral $CP$-odd and four charged states which each mix among themselves to form the mass eigenstates $h_i$, $A_i$ and $H^\pm_i$. 
In the following discussion, we will denote the lightest mostly electroweak Higgs state as $h$ and the lightest mostly right-doublet Higgs as $h_R$.
In the limit $\tanbetaR \to 1$, $h_R$ becomes massless at the tree level. In this case, the $SU(2)_R$- and the 
electroweak Higgs states decouple from each other and the second-lightest Higgs is purely $SU(2)_L$-doublet-like. 
The tree-level 
mass of $h$ is enhanced with respect to the MSSM prediction due to the effect of the extra $D$-
terms from the enlarged gauge sector. 
The absolute upper bound on this mass can be evaluated in the limits $\tanbetaR \to 1,\tanbeta \to \infty, \tan \beta_u \to \infty$ and is given by
\begin{equation}
m^2_{h,{\rm tree}}\big|^{\tanbetaR \to 1} \le \frac{1}{4} (g_L^2 + g_R^2 )\, v_L^2 \,,
\end{equation}
which is the generic upper limit for supersymmetric left-right theories where electroweak symmetry is broken by bidoublets \cite{Huitu:1997rr,Babu:2014vba} as well as in model variants where only the subgroup $U(1)_R\times U(1)_{B-L}$ survives
down to the TeV scale \cite{Krauss:2013jva,Hirsch:2011hg}.

\begin{figure}
\centering
\includegraphics[width=\linewidth]{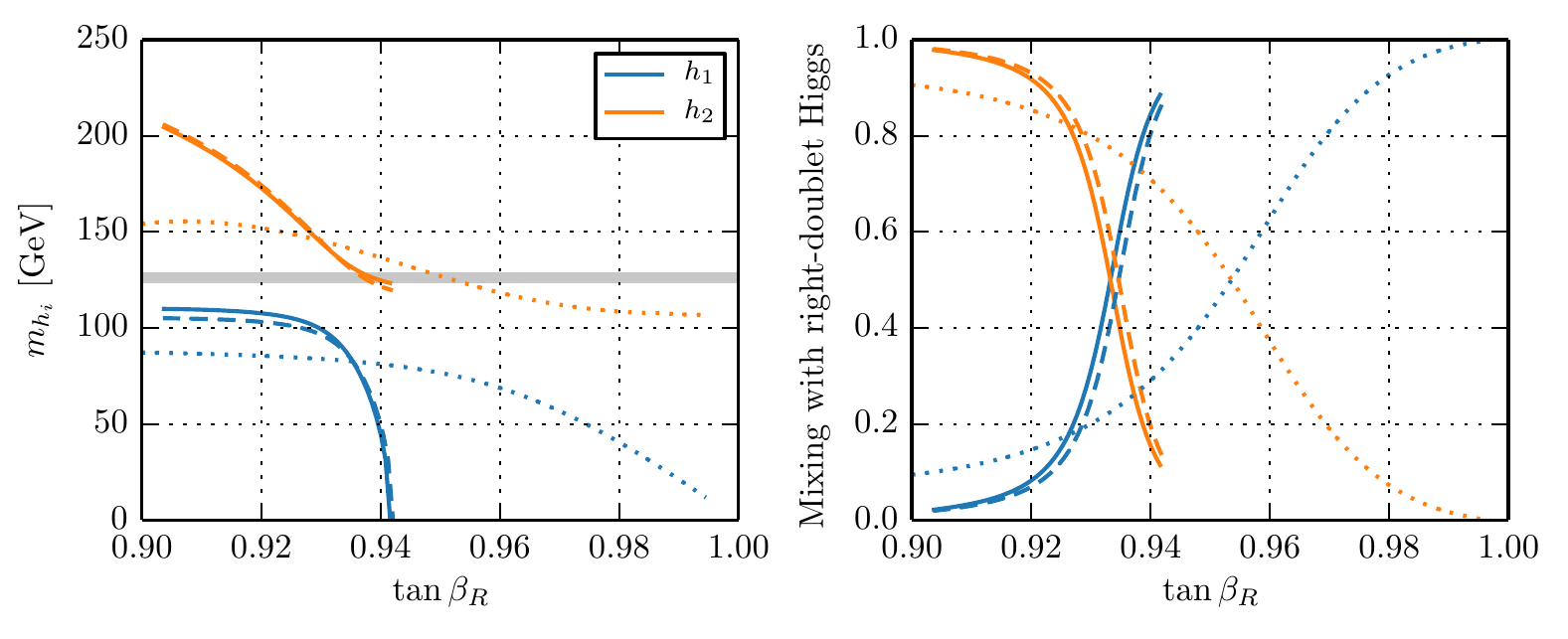}
\caption{Masses of the two lightest Higgs states as a function of $\tan \beta_R$. The results are shown at the tree level (dotted) as well as 
at the one/two loop level (dashed/solid lines). The grey band depicts the approximate mass required for a SM-like Higgs. The remaining parameters have been fixed to 
$m_0=M_{1/2}=\SI{1.2}{\TeV}$, $A_0 =\SI{1}{\TeV}$,  $\mu_\Phi^{(2,2)}=-\SI{2}{\TeV}$, $v_R=\SI{7}{\TeV}$, $\tanbeta=15$, $\tanbetau=10$, $M_\delta=\SI{1}{\TeV}$, $Y_{\delta_d}^i=0.09$.}
\label{fig:mh_loop}
\end{figure}

As soon as $\tanbetaR$ departs from one, a mixing between $h$ and $h_R$ sets in which rises with increasing $\Delta = 1- \tanbetaR$. This mixing also pushes up the heavier mass of both eigenstates.
Treating $\Delta$ as a small perturbation, one can evaluate the corresponding $2 \times 2$ mass matrix of said states which reads in the basis $(h,h_R)$:
\begin{align}
m^2_{h,h_R} &= \begin{pmatrix}
\frac{v_L^2 (D \, (g_L^2+g_R^2)-g_R^4 v_R^2)}{4 D} & \frac{- m^2_{A_R} \Delta\,  g_R^2 v_L v_R}{D} \\
\frac{-m^2_{A_R} \Delta \, g_R^2 v_L v_R}{D} & m^2_{A_R} \Delta^2
\end{pmatrix}\,,
\label{eq:simplified_2by2_higgsmatrix}
\end{align}
where $D\simeq 4(m^2_{A_R}+M_{Z^\prime}^2)$ and $m^2_{A_R} = -2\, B_{\mu_{\chi_c}}/\sin 2 \beta_R \simeq -2\, B_{\mu_{\chi_c}}$ is the mass of the pseudoscalar Higgs boson of the $SU(2)_R$ sector. 
After the level crossing of the eigenstates, 
$h_R$ continues getting more massive whereas the mass of
the electroweak eigenstate converges towards
\begin{equation}
m^2_{h,{\rm tree}}\big|^{\tanbetaR \to 0} \le \frac{1}{4} \left(g_L^2 + \frac{g_{BL}^2 g_R^2}{g_{BL}^2+g_R^2} \right)  v_L^2 = M^2_Z\,,
\end{equation}
which is exactly the same as in the MSSM. The last equality follows because of the relation between the hypercharge coupling $g_Y$ and the `new' couplings: 
$1/g_Y^2 = 1/g_{BL}^2 + 1/g_{R}^2$.

Taking into account the measured Higgs properties, the mixing between the Higgs states of the different $SU(2)$ sectors 
has to be small. Hence, there are two possibilities in our model: 
\begin{itemize}
\item values of $\tanbetaR$ close to one, resulting in a light $SU(2)_R$-doublet Higgs and a second-lightest Higgs 
with SM properties and an enhanced tree-level mass
\item significant departure from $\tanbetaR = 1$, in which case the lightest Higgs has SM properties but no $D$-term
 enhancement of the tree-level mass with respect to the MSSM.
\end{itemize}

In \cref{fig:mh_loop}, we show the masses as well as admixtures of the two lightest $CP$-even Higgs states at the tree level as well as the  one- and two-loop level as a function of $\tan \beta_R$.
Apart from the usual large corrections of several ten per-cent for the SM-like Higgs, the most apparent feature in the loop corrections
is the dependence on $\tan \beta_R$ which is altered at the loop level due to the coupling of $\chi_c$ to the vector-like coloured sector via $Y_{\delta_d}$: Since the average of the scalar masses can be smaller than the corresponding fermion mass, the loop corrections are negative in contrast to the well known feature of large positive (s)quark corrections in the MSSM. In  \cref{fig:mh_loop} we have chosen $M_{\delta}=\SI{1}{\TeV}$ as well as a relatively large coupling $Y_{\delta_d}=0.09$ (corresponding to $Y_{\delta_d}=0.26$ at $M_{\rm SUSY}$) to maximise these corrections.

As a consequence of those radiative corrections, a second-lightest SM-like Higgs can be accompanied by a very light $h_R$ state of $\mathcal O(10~{\rm GeV})$, in contrast to the constrained $U(1)_R \times U(1)_{B-L}$ model where the loop corrections in the absence of vector-like states always enhance $m_{h_R}$, i.e. one finds 
$\mathcal O(50~{\GeV})$ even for $\tan \beta_R \to 1$ \cite{Hirsch:2011hg,Hirsch:2012kv}. We remark that the branching ratio for the decay $h_2\to h_1 h_1$ is below a percent for these points even when the $h-h_R$ mixing is of $\mathcal{O}(\SI{10}{\%})$.

\subsection{Squark sector}
The down-squark mass matrix is enlarged to an $8\times 8$ matrix. The additional entries correspond to the vector-like quarks' scalar superpartners. The addition of these vector-like squarks modifies the expected hierarchy of the light--squark masses in comparison to the MSSM. Namely, we observe that the lightest down-squark is generically lighter than the lightest up-squark which is always the light stop $\tilde{t}_1$. This behaviour arises as the vector-like quarks modify the RGE running of the quark soft-masses, and have a potentially large mixing with the standard down-type squarks. 

To illustrate this behaviour we consider for the moment only the third-generation of left and right down-type squarks as well as the vector-like squarks. In the basis $\{ \tilde{b}_L,\tilde{b}_R, \tilde{\delta}_d, \tilde{\bar{\delta}}_d \}$ the mass matrix reads
\begin{align}\label{sbottom_massmatrix}
m_{\tilde{b},\tilde{\delta}}^2 &\simeq \begin{pmatrix} 
\left(m_Q^{(3,3)}\right)^2 & 0 & 0 & 0 \\
0 & \left(m_{Q_c}^{(3,3)}\right)^2 +\frac{1}{4} v_R^2 |Y^{(3)}_{\delta_d}|^2 & -\frac{1}{2}v_R\left(T^{(3)}_{Y_{\delta_d}} + \mu_{\chi_c} Y^{(3)}_{\delta_d} \right) & -\frac{1}{2} v_R M_\delta Y^{(3)}_{\delta_d} \\
0 &  -\frac{1}{2}v_R\left(T^{(3)}_{Y_{\delta_d}} + \mu_{\chi_c} Y^{(3)}_{\delta_d} \right) & |M_\delta|^2 + m_{\delta_d}^2 +\frac{1}{4} v_R^2 |Y^{(3)}_{\delta_d}|^2 & 0 \\
0 &  -\frac{1}{2} v_R M_\delta Y^{(3)}_{\delta_d} & 0 & |M_\delta|^2 + m_{\bar{\delta}_d}^2 
\end{pmatrix}\,.
\end{align}
Here, the electroweak VEVs have been neglected and we have assumed $\tanbetaR \to 1$ as these quantities give only a shift to the diagonal elements, but play a negligible role in the mixing with the vector-like states. From the form of the mass matrix we arrive at the following conclusions:
\begin{itemize}
\item There is no mixing between the left-sbottoms and the vector-like squarks based on these assumptions. 
\item For fixed values of $M_\delta$, the relative size of the mixing between the right-sbottoms and the vector-like states is determined by three parameters, namely $Y_{\delta_d}$, $A_0$ and $v_R$. Typically one requires these parameters to take large values in order to arrive at a phenomenologically viable model\footnote{Here we refer to the constraint that $Y_{\delta_d}$ must be sufficiently large to allow for spontaneous $SU(2)_R$ symmetry breaking and $v_R$ must be of order of several TeV to produce a sufficiently heavy $W'$.}. Therefore the right-sbottoms are typically strongly mixed with the vector-like states. This mixing reduces their mass compared to pure $\tilde{b}_{L/R}$ eigenstates.   
\end{itemize}
\begin{figure}
\centering
  \includegraphics[width=\linewidth]{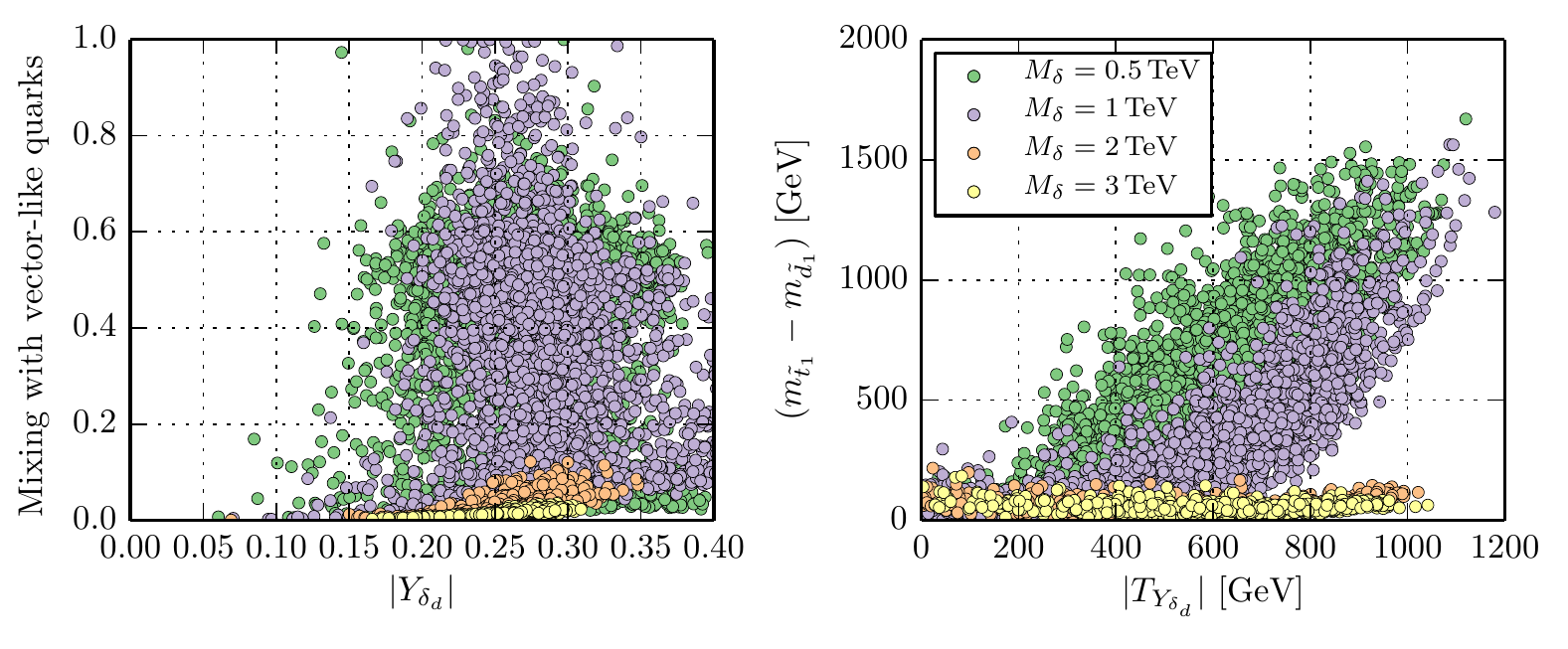}
\caption{The mixing of the lightest down-type squarks (left) and the splitting of both the lightest stop and down-squark masses (right) as functions of $Y_{\delta_d}$ and $T_{Y_{\delta_d}}$ at $M_{\rm SUSY}$ respectively. Here all input parameters are scanned over randomly for fixed values of $M_\delta$. The ranges of the parameters scanned over at the GUT scale are $v_R \in [6.5,9]\SI{}{\TeV}$, $\tanbeta, \tanbetau \in [1,30]$, $\tanbetaR \in [0.8,1]$, $m_0,M_{1/2} \in [200,2000]\SI{}{\GeV}$, $A_0 \in [0,3]\SI{}{\TeV}$, $\mu_\Phi^{(2,2)}\in [-3,3]\SI{}{\TeV}$ and $Y_{\delta_d} \in [-0.15,0.15]$.
}
\label{fig:Squark_Plot}
\end{figure}
In \cref{fig:Squark_Plot} on the left-hand panel the mixing of the lightest sbottom with the vector-like states is shown as a function of $Y_{\delta_d}$, where a value of 1.0 corresponds to a purely vector-like squark and zero corresponds to a pure MSSM sbottom state. Here we observe that depending on $M_\delta$ there exists a minimum value of $Y_{\delta_d}$ required for significant mixing with the vector-like states. In the right-hand panel we show the effect of $T_{Y_{\delta_d}}$ on the splitting of the squark masses. As \cref{sbottom_massmatrix} suggests, $T_{Y_{\delta_d}}$ contributes strongly to this splitting. One should note that the value of this trilinear coupling is strongly correlated with $M_{1/2}$ due to RGE running, increasing with larger $M_{1/2}$.

\begin{figure}
\centering
  \includegraphics{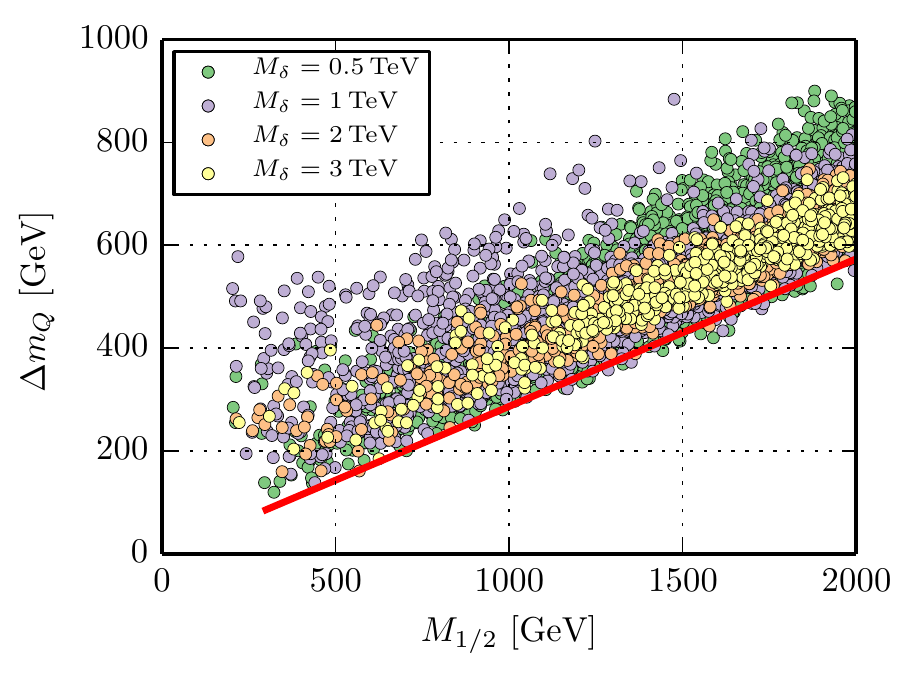}
\caption{A random scan which illustrates the splitting of the squark soft-masses. Also shown in red is the analytic expression based on the asymmetry of the left-right gaugino mass terms. Here all input parameters are scanned over randomly for fixed values of $M_\delta$. The ranges of the parameters scanned over at the GUT scale are $v_R \in [6.5,9]\SI{}{\TeV}$, $\tanbeta, \tanbetau \in [1,30]$, $\tanbetaR \in [0.8,1]$, $m_0,M_{1/2} \in [200,2000]\SI{}{\GeV}$, $A_0 \in [0,3]\SI{}{\TeV}$, $\mu_\Phi^{(2,2)}\in [-3,3]\SI{}{\TeV}$ and $Y_{\delta_d} \in [-0.15,0.15]$. 
}
\label{fig:Squark_Plot_M12}
\end{figure}

RGE running effects result in a splitting of the quark soft masses where $(m_Q^{(3,3)})^2 > (m_{Q_c}^{(3,3)})^2$. This splitting arises through two main sources. Firstly, the running of the gaugino masses in the left-right sector is asymmetric. This results in the splitting being a function of $M_{1/2}$ which can be analytically estimated at the one-loop level:
\begin{align}
\Delta m_Q^2 &\equiv (m_Q^{(3,3)})^2 - (m_{Q_c}^{(3,3)})^2 \notag\\
&\simeq \frac{M_{1/2}^2}{4} \left[1+16\pi^4\left(\frac{8}{\left(8\pi^2-3 g_{\rm GUT}^2 \ln \left(\frac{M_{\rm SUSY}}{M_{\rm GUT}}\right)\right)^2} - \frac{3}{\left(-4\pi^2+ g_{\rm GUT}^2 \ln \left(\frac{M_{\rm SUSY}}{M_{\rm GUT}}\right)\right)^2}\right)\right]\notag \\
&\simeq \SI{8.2E-2}{} M_{1/2}^2
\end{align}
Here, $g_{\rm GUT} \simeq 0.8$, $M_{\rm GUT} \simeq \SI{1.5E17}{\GeV}$ and $M_{\rm SUSY}\simeq \SI{2.5}{\TeV}$. In \cref{fig:Squark_Plot_M12} the results of the fully numerical scan are shown. The bold red line corresponds to the above function, whereby we see that this function provides an adequate approximation to the minimal splitting of the soft-masses. Secondly, as is also illustrated by \cref{fig:Squark_Plot_M12}, additional splitting occurs due to $Y_{\delta_d} \neq 0$. The precise value of this contribution depends strongly upon numerous parameters in the model. 

\subsection{Comments on recent LHC excesses}

\subsubsection{ATAS diboson excess} \label{sec:diboson}

The ATLAS collaboration observed an excess with a local significance of $3.4~\sigma$ at an invariant mass of around 2~TeV in their search for diboson events in the $W\,Z$ channel \cite{Aad:2015owa}. Many models have been proposed to explain that excess, in particular left-right symmetric models
where a $W'$ decays into $W\,Z$ caused by a significant $W-W'$ mixing. 
The best-fit point for the amount of mixing that needs to be present in minimal non-SUSY left-right models lies around $ \Theta_{WW'} \simeq \SI{1.4E-3}{}$, while $g_R/g_L \simeq 0.6$ \cite{Brehmer:2015cia}. 
The ratio $g_R/g_L$ cannot be much larger than this value 
in the minimal model as the bounds from dijet searches are otherwise too constraining \cite{Khachatryan:2015sja}. In our model $g_R$ is not a free parameters, but predicted to be $g_R/g_L \simeq 0.9$. This is still consistent with observation because of the extra decay modes of $W'$ into supersymmetric particles which increase its width and accordingly reduce its dijet production cross section.
Parameterising the process in the narrow width approximation as $\sigma (pp \to W')\times {\rm BR}(W' \to W\,Z)$, 
with $\sigma(pp \to W') \propto g_R^2$ and to a first approximation BR($W' \to W\,Z$)$= \Gamma_{WZ}/\Gamma_{\rm tot} \propto  \Theta_{WW'}^2/(\xi\, g_R^2)$, where $\xi>1$ parameterises the increased width due to the supersymmetric final states, 
we see that, to first order, $g_R$ drops out of those considerations and we actually need somewhat larger values of $\Theta_{WW'}$ than in the simplest model due to the influence of $\xi$.

Using \cref{eq:wwp_mixing}, we see that $\Theta_{WW'}$ is maximised for $\tanbeta =1 $ and $\tanbetau = \tanbetad $ reaching the required size of $\Theta_{WW'} \simeq \SI{1.4E-3}{}$. However, this suffers from several problems: the former condition corresponds to a saddle point of the potential and predicts a tiny tree-level Higgs mass, while the latter condition leads to non-perturbative Yukawa couplings, cf. \cref{eq:yukawadef2}, which makes it impossible to achieve the required size of $W-W'$ mixing. 
 
Thus far we have neglected the impact of the extended (s)quark sector and the effect of mixing with the vector-like states. In Ref.~\cite{Collins:2015wua}, the Ansatz was made that the light quarks are predominantly comprised by extra vector-like quark states which considerably reduces the $W'$ coupling to light quarks. The excess can then be fit with a $W'-\bar{q}'-q$  coupling smaller by a factor of 5 compared to the SM value and $\tan \beta < 5$ for $g_R/g_L \simeq 0.9$. 
 The latter is still clearly disfavoured in light of Higgs data keeping in mind the limit on the mass spectrum coming from gauge coupling unification, whereas the huge mixing is very hard to realize:
casting aside all problems of inducing large flavour changing neutral currents and just focusing on the feasibility of this possibility in our model, we would need a hierarchy of $v_{\bar \chi_c} Y_{\delta_d}/(\sqrt{2}\, M_{\delta_d}) \sim 5$ , which is in principle achievable for $M_{\delta_d}$ of $\mathcal O(100~{\rm GeV})$. 

However, the problem is expected to occur in the down-squark mixing where the absence of a reasonably large $M_{\delta_d}$ typically leads to tachyonic states.
In order to easily see this, we apply a few approximations. The soft SUSY-breaking masses $m^2_{Q_c}$ and $m^2_{\delta_d}$ will be of similar size even after RGE running as the 
 contributions from the gluino are the same. Therefore we will set both to the common squared mass $m^2$. Furthermore, if we write $T_{\delta_d}$ as $A_0 Y_{\delta_d}$ and consider the limit $M_{\delta_d} \to 0$, the diagonalization of \cref{sbottom_massmatrix} leads to the mass 
 eigenstates 
 \begin{align}
  m^2 + \frac{1}{4} v_R^2 |Y_{\delta_d}|^2 \pm \frac{1}{2} Y_{\delta_d}\, v_R (A_0 + \mu_{\chi_c})\,,
 \end{align}
which leads to a negative state for typical GUT-scale values of $m_0,\, A_0$ because of the strong running of $A_0 Y_{\delta_d}$ and accordingly $A_0 \gg \mu_{\chi_c}$ at $M_{\rm SUSY}$ unless the value of $A_0$ at the GUT scale is tuned to approximately cancel $\mu_{\chi_c}$.
In light of the numerous essential tunings required to obtain the necessary conditions, we conclude that our model cannot fit the excess and simultaneously remain a consistent high-scale model. 

\subsubsection{CMS $eejj$ excess}

In the seach for dilepton plus dijet events in the 8~TeV data, CMS has announced an excess of signal events with a 
local significance of 2.8~$\sigma$ \cite{Khachatryan:2014dka}.
This excess can be interpreted as the on-shell
 production of a $W_R$ with $M_{W_R} \simeq 2.1~$TeV and its subsequent decay into $\nu_h \, \ell$, with 
 $\nu_h \to \ell j j$. 
 The comparatively low $pp \to \ell \ell j j$ cross-section which fits the excess 
 cannot be explained in the framework of simplified left-right-symmetric models which only introduce a $W'$ and three copies of right-handed neutrinos, predicting a cross-section which is higher by a factor of $3 - 4$ if $g_R \simeq g_L$.
 Furthermore, as the excess features dileptons of differing signs, it cannot be explained by 
 heavy neutrinos of Majorana nature which would predict $\ell^\pm \ell^\pm$ final states at the same rate 
 as $\ell^\pm \ell^\mp$. 
 As discussed in section~\ref{subsec:lepton}, the heavy neutrinos in the present model form quasi-Dirac pairs because of the inverse seesaw 
 mechanism at work. As a consequence, the lepton appearing in the $\nu_h$ decay will have the opposite sign as the one in the 
 $W'$ decay to $\nu_h \ell$, in agreement with the measured effect.
 Additionally, a reduction of the cross section is achieved by the 
 interplay with the lightest charged Higgs state $H^\pm_1$, analogously to Ref.~\cite{Krauss:2015nba}: the 
mass of the mostly $H_{d}^{1-}$-like state is, to a good approximation, $m_{H^\pm_1}^2 \simeq \frac{g_R^2}{4} (v_{\bar \chi_c}^2-v_{\chi_c}^2)$, which is usually in the range of a few hundred GeV for the typical values of $\tan \beta_R$.
Heavy neutrinos couple to this state by the admixture of $\chi_c^-$. Consequently, with $m_{\nu_h}$ of $\mathcal O(1~{\rm TeV})$, 
the two-body decays $\nu_h \to \ell\, H^\pm_1$  reduce the branching ratio of $\nu_h \to \ell j j $ by several 
tens of percent and hence also the $\ell \ell j j$ cross section by the right amount. 

A further reduction is possible due to the mixing 
of the light neutrinos with the heavy states which opens the additional decay modes $\nu_h \to \ell W/ \nu h/\nu Z$. 
 As the mixing only depends on the ratio $\frac{m_D}{v_{\chi_c}} Y_S^{-1}$ but the light neutrino masses scale with $\mu_S \, m_D^2$, neutrino data also allows the 
possibility that $\mu_S$ is very small and $m_D$ sizeable. Hence, while the effect of the mixing is still small for values of $\mu_S$ of $\mathcal O(10^{-4}~{\rm GeV})$, it already gets important for $\mu_S=10^{-5}~$GeV, where for the required masses of $M_{W'}\simeq 2\,m_{\nu_R} \simeq 2~$TeV those decays are already of the same size as the $W'$-mediated three-body decay.
Should the excess be confirmed, this could be the main source of cross section reduction in the case of a large deviation of $\tan \beta_R$ from one and the associated heavier charged Higgs state.


\section{Conclusion}
\label{sec:conclusion}
We have presented a constrained left-right supersymmetric model which predicts  a low $SU(2)_R\times U(1)_{B-L}$ breaking scale. The model is constructed in a manner where gauge coupling unification is maintained, based on the requirement that $SU(2)_R\times U(1)_{B-L}$ is broken purely through $SU(2)_R$ doublets. As the left-right breaking scale is assumed to be close to the SUSY scale, gauge coupling unification dictates that additional matter must be introduced. This extra matter takes the form of vector-like quarks and leptons charged under $U(1)_{B-L}$ but being
singlets with respect to the $SU(2)$ factors. 

Due to the fast running of the $U(1)_{B-L}$ gauge coupling and large one-loop threshold corrections, the model predicts a unification scale close to the string scale. These large threshold corrections are a product of large values of the gauge coupling beta functions in conjunction with a large spread in the mass spectrum. For unification to remain unspoilt by threshold corrections, one naturally predicts the $SU(2)_R$ breaking scale to lie close to $M_{\rm SUSY}$.  Finally, the presence of vector-like quarks are an essential ingredient in driving spontaneous symmetry breaking in the left-right phase: under the assumption of mSUGRA-like boundary conditions, the couplings of these quarks must be non-vanishing to trigger radiative gauge symmetry breaking. 

We have demonstrated why the usual paradigm of using vector-like quarks in conjunction with the seesaw mechanism provides insufficient degrees of freedom to fit both the quark masses and mixings simultaneously. Subsequently, we have implemented both the quark masses and mixing through the introduction of an additional Higgs bi-doublet, raising the total number of electroweak VEVs to four. 

The phenomenology of this model contains a number of interesting features. Here, we have focused on the mass spectrum. Firstly the CP-even Higgs sector displays two distinctive tendencies. For $\tanbetaR \to 1$, the lightest CP-even Higgs mass tends to $\mathcal{O}(10)\SI{}{\GeV}$ values while the second lightest Higgs becomes SM-like. For sufficient deviation from $\tanbetaR =1$, the MSSM-like limit is produced. Note, that the
lightest state is essentially a SM gauge singlet. In the squark sector due to both the RGE running and the enlarged down-squark sector the lightest down squark is always lighter than the lightest stop. 

Finally we comment on two of the recently observed excesses seen at the LHC. We show that due to constraints arising from RGE running sufficient mixing of the $W$-bosons cannot be achieved to naturally explain the diboson excess. Even if one can make use of the vector-like quarks to weaken this conclusion, we consider a concrete realisation of that idea in this model to be very unlikely. However, the CMS $eejj$ excess can be successfully explained due to the model incorporating an inverse seesaw and the presence of extra two-body decays.

\section*{Acknowledgements}

M.E.K.\ and W.P.\ have been supported by  the DFG, project nr.\ PO-1337/3-1.
M.H. is supported by the Spanish MICINN grants FPA2014-58183-P and Multidark
CSD2009-00064 (MINECO), and PROMETEOII/2014/084 (Generalitat
Valenciana). M.E.K. is supported by the BMBF grant 00160287.


\providecommand{\href}[2]{#2}\begingroup\raggedright\endgroup

\end{document}